\documentclass[aps,pra,twocolumn,bibnotes,showpacs]{revtex4}
\usepackage{amsmath,amsfonts,amssymb,amsbsy,graphicx,bbm,times,hyperref}
\usepackage{color}

\newcommand{\be}{\begin{equation}}
\newcommand{\ee}{\end{equation}}
\newcommand{\bea}{\begin{eqnarray}}
\newcommand{\eea}{\end{eqnarray}}
\newcommand{\ket}[1]{|#1\rangle}
\newcommand{\bra}[1]{\langle#1|}

\newcommand{\eq}[1]{Eq.~(\ref{#1})}

\begin{document}

\title{Simple proof of the robustness of Gaussian entanglement in bosonic noisy channels}

\author{Gerardo Adesso}
\affiliation{School of Mathematical Sciences, University of Nottingham, University Park, Nottingham NG7 2RD, United Kingdom}

\date{September 29, 2010}

\begin{abstract}
The extremality of Gaussian states is exploited to show that  Gaussian states are the most robust, among all possible bipartite continuous-variable states at fixed energy, against disentanglement due to noisy evolutions in Markovian Gaussian channels involving dissipation and thermal hopping. This proves a conjecture raised recently in [M. Allegra, P. Giorda, and M. G. A. Paris, Phys. Rev. Lett. {\bf 105}, 100503 (2010)],  providing a rigorous validation of the conclusions of that work.
The problem of identifying continuous variable states with maximum resilience to entanglement damping in more general bosonic open system dynamical evolutions, possibly including correlated noise and non-Markovian effects, remains open.
\end{abstract}


\maketitle

Continuous-variable (CV) systems such as light modes and ultracold atomic ensembles \cite{COVAQIAL} provide advantageous resources to achieve unconditional implementations of quantum information processing \cite{brareview}, ranging from teleportation protocols \cite{Furusawa98} to quantum key distribution \cite{cvcrypto} and one-way quantum computation \cite{1waycv}. Gaussian states  and Gaussian operations, that represent respectively the most natural and easily controllable light states as well as the set of manipulations efficiently realizable by linear optics, have traditionally occupied a privileged role in all such implementations. Furthermore, by virtue of their mathematical simplicity compared to general states living in infinite-dimensional Hilbert spaces, bosonic Gaussian states have been and are the preferred testgrounds for a broad variety of investigations into the structure, nature, and dynamics of CV entanglement and quantum correlations \cite{ourreview}.

Some prominent limitations of a Gaussian-only toolbox have been however recently exposed in several contexts \cite{nogos}, stimulating vigorous theoretical and experimental research into the realm of non-Gaussian state engineering and characterization \cite{belliniecc}, to assess and harness the potentially enhanced performance of de-Gaussified CV resources for quantum teleportation \cite{degausstelep}, entanglement distillation \cite{distillscheme}, parameter estimation \cite{souza}, universal quantum computation \cite{cvcomput,1waycv}, non-locality tests \cite{loopholefree}, etc.
Still, one of the most powerful features of Gaussian quantum states is their {\it extremality} \cite{extra} in the space of all CV states, that allows to formulate valuable bounds on suitable entanglement measures and entropic degrees for a general non-Gaussian state $\varrho$, based on the corresponding (easier to compute) properties of the Gaussian state $\sigma$ with the same first and second statistical moments as $\varrho$. This has important consequences for the security of CV quantum key distribution \cite{qkdextra}.

The transmission of one-mode and multi-mode, possibly entangled beams between distant locations, a basic necessity for the  realization of a distributed quantum communication network \cite{qinternet}, is unavoidably affected by various types of noise. While phase diffusion (dephasing) dampens the coherences in the Fock basis transforming Gaussian states into non-Gaussian ones, the exposition to dissipative losses and thermal hopping yields an instance of a Gaussian channel, that preserves the Gaussian character of the states sent through it.

Recently, Allegra, Giorda and Paris have attempted a comparison (by means of several separability criteria) of the decoherence effects due to the latter types of channels, on special families of Gaussian as well as non-Gaussian two-mode entangled states, all initially having the same mean energy \cite{allegra}. Among the main findings of their work, they formulate a conjecture on the ``robustness of Gaussian entanglement'', according to which initial Gaussian pure states should be the ones whose entanglement is the last to vanish, compared to other classes of initial non-Gaussian pure states with equal energy, when they are all evolving through a noisy Markovian  channel with loss and thermal hopping.

In this Brief Report, we prove the conjecture of Ref.~\cite{allegra} by resorting to the extremality of Gaussian states \cite{extra}, in particular to the fact that Gaussian states maximize entanglement among all CV states with fixed energy \cite{vanenk}. Our simple result establishes in full generality that  no advantage can be gained by encoding information into entangled non-Gaussian states when they have to be transmitted through Gaussian noisy  channels (under dissipation and/or amplification), and that Gaussian states are, hence,  the most robust CV states indeed against photon losses and thermal hopping. An outlook on the  problem of  identification of the most robust CV entangled states for more general bosonic noisy evolutions, including dephasing channels, is provided in detail.

We consider general pure two-mode CV states $\varrho_{12}(0)$ with fixed initial mean energy $\bar{n}_0 = \mbox{Tr} [\varrho_{12}(0) (a^\dagger_1 a_1 + a^\dagger_2 a_2)]$, where $a_j$, $a^\dagger_j$, are the ladder operators of mode $j=1,2$ satisfying the canonical commutation relations $[a_i,a^\dagger_j]=\delta_{ij}$. Among all the possible input states, the two-mode squeezed state $\varrho_{12}^G(0)=\ket{\psi}\bra{\psi}$, where $|\psi\rangle=\sum_n{\lambda^n\sqrt{1-\lambda^2}}\ket{n,n}$ and $\bar{n}_0 = \lambda^2/(1-\lambda^2)$, is the only Gaussian instance (up to local unitary operations).
We let each possible initial state undergo a dissipative evolution through the channel described by the following master equation \cite{allegra},
\begin{equation}\label{ME}
\dot \varrho_{12}(t)  =  \sum_{j=1,2}\frac{\Gamma}{2}N_{j} \: L[a_{j}^{\dag}]\varrho_{12}(t)
+\frac{\Gamma}{2}(N_{j}+1)\:L[a_{j}]\varrho_{12}(t)\,,
\end{equation}
that encompasses losses and thermal hopping in the presence of nonclassically fluctuating local environments. The dot denotes time-derivative and the Lindblad
superoperator is defined as $L[O]\varrho =  2 O\varrho O^{\dag} - O^{\dag}
O\varrho -\varrho O^{\dag} O$. Here $\Gamma$ is a loss coefficient and $N_{j}$ is the mean
photon number in the stationary (thermal) state of each mode.
The master equation (\ref{ME})  admits the operator solution \cite{allegra,gmd94}:
$\varrho_{12}(t) = \Lambda_t \varrho_{12}(0) = \mbox{Tr}_{34} [U_t (\varrho(0)
\otimes \nu_{34} ) U_t]$, where $\Lambda_t$ denotes the evolution map corresponding
to the noisy channel; $3, 4 $ are two additional fictitious modes in a thermal
state $\nu_{34} = \nu_3 \otimes \nu_4$;  $U_t=U_{13}(\zeta_t) \otimes
U_{24}(\zeta_t)$ and $U_{ij}(\zeta_t)= \exp(\zeta_t a_i^\dag a_j -
\zeta_t^* a_j^\dag a_i)$ amounts to a beam-splitter transformation
with $\zeta_t=\arctan(e^{\Gamma t}-1)^{1/2}$.

\begin{figure}[t]
\includegraphics[width=8.5cm]{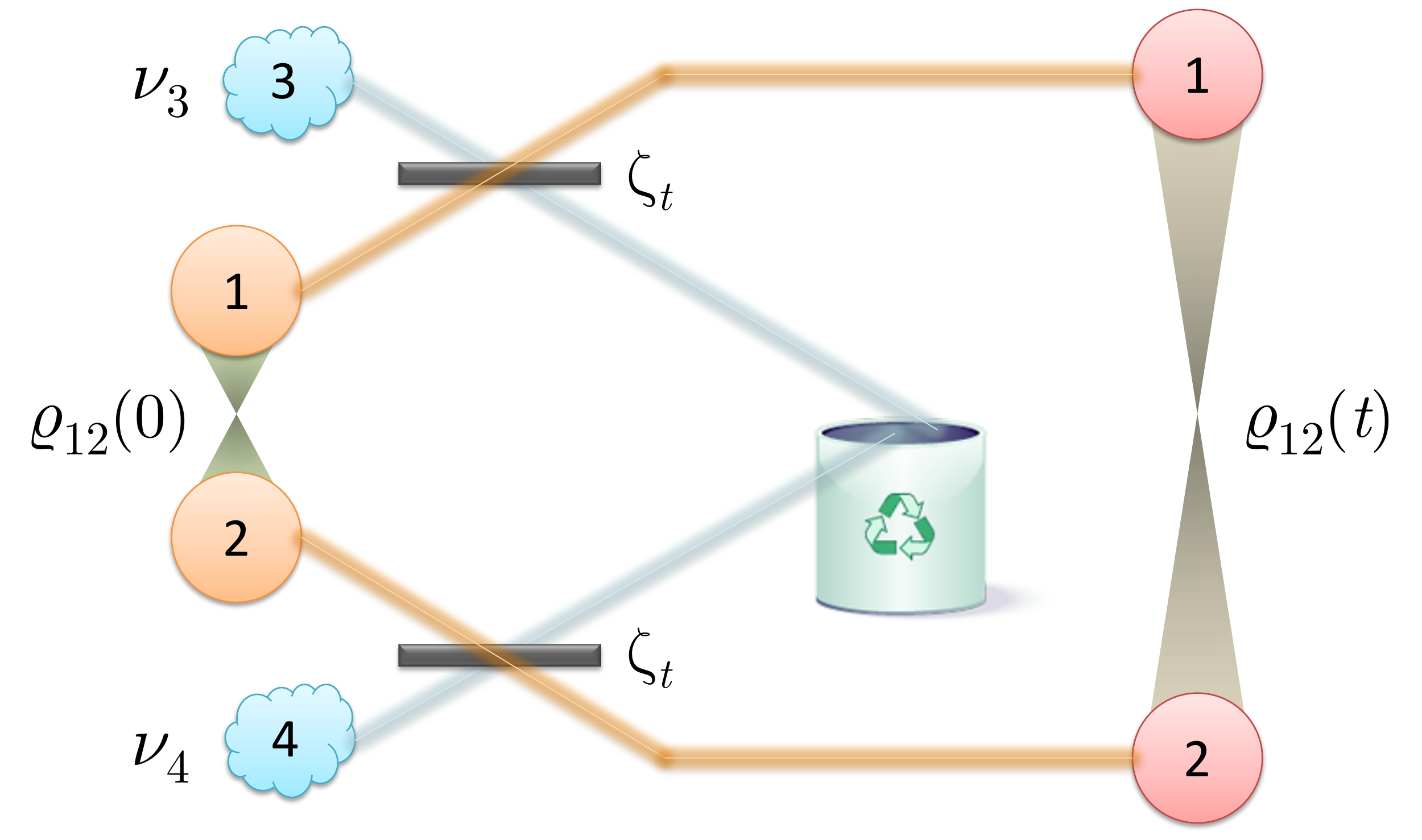}
\caption{(Color online) Scheme of the channel in \eq{ME}.}\label{figdiag}
\end{figure}

The action of the noisy evolution in \eq{ME} can be thus simply modeled as letting each mode interfere with an ancillary thermal mode via a beam-splitter  with transmissivity $\cos \zeta_t$, and eventually tracing over the ancillae, as depicted schematically in Fig.~\ref{figdiag}. In this notation, $N_1$ ($N_2$) is the mean photon number of the ancillary mode $\nu_3$ ($\nu_4$).
In the Heisenberg picture, the beam-splitter transformation acts on the mode operators as follows \cite{nielsen}
\begin{equation}\label{bsheis}
a_j^{(t)} = a_j \cos \zeta_t  + a_{j+2} \sin \zeta_t\,,
\end{equation}
where $a_{3,4}$ are the annihilation operators of the ancillary modes.
We can calculate the mean energy of the generic state $\varrho_{12}(t)$ at any time $t$ during the evolution:
\begin{eqnarray}
\bar{n}_t &=& \mbox{Tr} \left[(\varrho_{12}(0) \otimes \nu_{34}) \big({a^{(t)}_1}^\dagger a^{(t)}_1 + {a^{(t)}_2}^\dagger {a^{(t)}_2}\big)\right]  \nonumber \\
          &=& \bar{n}_0 \cos^2 \zeta_t + (N_1+N_2) \sin^2\zeta_t\,,
\end{eqnarray}
where all the additional cross-terms vanish because the ancillary thermal states have zero first moments, entailing $\langle a_{3,4}\rangle_{\nu_{34}}=\langle a^\dagger_{3,4}\rangle_{\nu_{34}}=0$.
What is important here, is that the energy of the evolving states $\varrho_{12}(t)$ at any time only depends on the initial energy and on the channel parameters, i.e., if a set of states $\varrho_{12}(0)$ enter the channel all with the same energy, as it is the case in \cite{allegra} and in the present discussion, their energies will remain equal to each other (yet globally affected by the open system dynamics) throughout the whole evolution, regardless of the specific form of the states and of their Gaussian or non-Gaussian nature. Let us also remark once more that the channel in \eq{ME} is a Gaussian channel, i.e. the state $\varrho^G_{12}(t)$ evolved from the two-mode squeezed state will remain Gaussian during the whole evolution. The final important ingredient we need is the extremality of Gaussian states. If one adopts a {\it bona fide} measure of entanglement ${\cal E}$ \cite{pv}, which is continuous and strongly superadditive, such as the distillable entanglement, or the squashed entanglement \cite{squashed}, then the Gaussian state $\varrho^G_{12}(t)$ with energy $\bar{n}_t$ is the {\it most entangled} among all possible CV states $\varrho_{12}(t)$ with the same energy \cite{extra,vanenk}, as a simple consequence of the maximum entropy property of Gaussian states \cite{holevo}.

Collecting all the above observations, we can conclude that ${\cal E}[\varrho^G_{12}(t)] \ge {\cal E}[\varrho_{12}(t)]$ at any time $t$, in other words the Gaussian two-mode squeezed state starts in advantage, in terms of entanglement, over any other CV state with the same energy, and maintains its advantage throughout the whole noisy evolution. Given that (as soon as $N_1,N_2 \neq 0$) the channel in \eq{ME} takes a  finite ``separation time'' $\tau[\varrho_{12}(0)]$ to erase  all the entanglement in any initial quantum state $\varrho_{12}(0)$ \cite{allegra}, the entanglement in the Gaussian state must be the last one to vanish, i.e., $\tau[\varrho^G_{12}(0)] \ge \tau[\varrho_{12}(0)]$ for all CV states $\varrho_{12}(0)$ with the same mean energy as $\varrho^G_{12}(0)$. This proves the robustness of Gaussian entanglement conjecture of Ref.~\cite{allegra}, for continuous and strongly superadditive bipartite entanglement monotones, in the general case of $N_1 \neq N_2$ and without the need to restrict $\varrho_{12}(0)$ to be in the form of a photon-number-entangled-state (i.e., a two-mode state whose one-mode marginal density matrices are diagonal in the Fock basis). Actually, being essentially a consequence of extremality, the present result holds for any set of, not necessarily pure, iso-energetic input states $\varrho(0)$ of an arbitrary number of modes  that include a Gaussian one $\varrho^G(0)$, and for general Markovian Gaussian channels.

All the speculations of Ref.~\cite{allegra} that were based on this conjecture can be now regarded as valid and truthful: Gaussian entanglement {\it is} extremal
in terms of robustness against decoherence due to noise
and dissipation. It is important to notice, however, that if entanglement is measured e.g.~by the (logarithmic) negativity \cite{pv}, which does not satisfy the hypotheses of the theorem in \cite{extra}, then the extremality of Gaussian states can be violated by tiny amounts \cite{vanenk,extra}, hence narrow regimes might be found where a particular non-Gaussian state could overcome $\varrho^G_{12}(t)$ in the entanglement hierarchy based on negativities. This is nonetheless to be regarded more as a pathological feature of the employed entanglement measure, rather than as a physically meaningful violation or counterexample to the robustness conjecture. Let us remark once more that our general proof in fact holds for any continuous and strongly superadditive bipartite entanglement monotone (such as the squashed entanglement \cite{squashed}). It becomes quite clear, henceforth, how the degree of non-Gaussianity  -- measured as distance from a Gaussian reference state with the same covariance matrix \cite{nonG} -- of different instances of $\varrho_{12}(0)$'s is set to play a key role in assessing their robustness to noise and their entanglement evolution, as evidenced by several examples in Ref.~\cite{allegra}.

Our analysis generalizes the one in Ref.~\cite{vanenk}, where the two-mode squeezed Gaussian state had already been identified as ``the most robust state of light'' against evolutions in purely dissipative channels [i.e.~with $N_1=N_2=0$ in \eq{ME}].
These results do not straightforwardly extend to more general scenarios in which other types of noisy channels are taken into account, that can violate Gaussianity and/or Markovianity. In these cases, the Gaussian nature of transmitted states is not preserved, and the state evolved from $\varrho^G_{12}(0)$ immediately loses its privileged role for $t>0$, entering the competition among the $\varrho_{12}(t)$'s as a peer.
This is the case, for instance, if purely local phase diffusive decoherence channels are considered, which are non-Gaussian channels modeled by a master equation of the form
\begin{equation}\label{MEdephas}
\dot \varrho_{12}(t)  =  \sum_{j=1,2}\frac{\Gamma}{2} \: L[a_{j}^{\dag}a_j]\varrho_{12}(t).
\end{equation}
 For such noise models, the coherences in an initial two-mode squeezed state are degraded undergoing a non-Gaussian evolution, eventually driving the two modes into an uncorrelated product of thermal states, each corresponding to the reduced one-mode density matrix of the input two-mode state. In this scenario, there is no apparent hierarchy regulating the dynamical comparison between Gaussian and non-Gaussian initial states $\varrho_{12}(0)$, when energy or alternative resources (i.e., the squeezing degree) are  fixed at time $t=0$. More complicate is the situation when a two-mode dephasing channel, with correlated noises affecting each mode, is taken into account; in that case, not even the marginal states are preserved.

To formulate an outlook, we feel that  even more challenging, yet surely important, would be to investigate the case of more realistic channels where dissipation, thermal hopping, as well as phase diffusion and possibly memory effects and correlated noise  may all simultaneously take place. Some of these mechanisms are currently investigated, for instance, in the context of unveiling quantum coherence effects and noise-assisted energy transport phenomena in biological systems \cite{caruso}.  The problem of entanglement evolution and robustness turns thus into a highly non-trivial arena, and since the powerful extremality results  \cite{extra} are of little or no use at all in these more general cases, a numerical approach along the lines of Ref.~\cite{allegra} should be the preferable avenue to pursue. This will be the subject of further work. The main result of this Brief Report should then be recovered in the limit of negligible dephasing, uncorrelated noises and memoryless channels, but we feel that no reliable prediction can be made at this stage about which type of state, in the uncountable arena of CV states, might gain the crown of the most robust state of light in the intermediate regime where all the diverse noise effects are comparable.

\smallskip

Fruitful discussions with Matteo Paris and Alex Monras are acknowledged.

\end{document}